\pdfoutput=1

\documentclass[prb,preprint]{revtex4-1} 

\usepackage{amsmath}  
\usepackage{amsfonts} 
\usepackage{graphicx} 
\usepackage{color}

\begin{document}


\title{On the electrostatic equilibrium of charges and cavities in a conductor}

\author{Aritro Pathak}
\email{ap323@brandeis.edu} 
\altaffiliation[permanent address: ]{12 East Road Jadavpur, Kolkata, India, Pin Code:700032} 
\affiliation{Department of Mathematics, Brandeis University, Waltham, MA 02453}


\date{\today}

\begin{abstract}
 We consider a charged conductor of arbitrary shape, in electrostatic equilibrium, with one or more cavities inside it, and with fixed charges placed outside the conductors and inside the cavities. The field inside a particular cavity is then only due to charges within that cavity itself and to the   surface charge induced on the surface of the same cavity. A similar statement holds for the exterior of the conductor. Although this is an elementary property of conductors, it is not a trivial statement, as explained in this article. Undergraduate texts in electrodynamics do not discuss at length or provide a complete argument for an important problem such as this. Two simple and complete proofs are provided in this note with the help of the standard electrostatic uniqueness theorems. 
 
\end{abstract}
\maketitle 

\section{INTRODUCTION}

In a widely used undergraduate text \cite{Griffith} on Electrodynamics by Griffiths, the case of a neutral spherical conductor at electrostatic equilibrium with an arbitrarily shaped cavity is considered. Inside the cavity there is a point charge $q$. 

The field outside the conductor depends only on the charge induced on the outside surface of the conductor. The field inside the cavity depends only on the charge $q$ and on the charge that gets induced on the inner surface of the cavity. Speaking in an informal way, charges are induced on the cavity surfaces in such a way that in equilibrium, the field due to the charges in one domain of space does not "penetrate" the conductor and enter a different domain of space. 

This is not an obvious result; there is no immediate reason for the charges to get distributed this way. The following physical motivation is plausible: during the initial transient stage, when the charges are getting distributed on the surfaces, the field due to all charges outside the conductor (given volume charge density as well as the induced surface charge density outside the conductor), at the location of the surface of the cavity, moves around the charges on the surface of cavity so that finally at equilibrium they have been redistributed to negate any residual field within the cavity, due to charges outside the conductor. Simultaneously, there is an analogous effect on the surface charge on the outside surface of the conductor as it also get redistributed by the field of all the charges in the cavity, so that finally the field in any particular domain is only due to the charges in that particular domain. 

Following the example in Griffiths's book\cite{Griffith} there is a discussion where it is stated that the Uniqueness Theorems that follow in the subsequent chapter would address the problem. While a plausible heuristic argument is presented, no exact argument is given. 

In this article, we generalise this problem to the more involved one with a charged conductor containing two cavities with volume charges inside each, and provide a rigorous and exact argument for why the surface charges are induced in this special way. The proof of the original problem follows from the proof of this more general problem. It illustrates that the Uniqueness Theorems are indispensable tools in formulating theoretical arguments in electrostatics. We note that because the result of the Uniqueness Theorems also holds for any region containing linear dielectric material, (with the dielectric constant not necessarily uniform or even continuous) our cavities or the outside of the conductor might as well be filled fully or partly with linear dielectric material.

To solve this problem, some are tempted to consider a particular cavity in isolation with it's own volume charges, where the boundary of the cavity is at the constant potential value of the conductor in the actual problem. The field within the cavity, and hence also the induced cavity surface charge distribution in this isolated problem, are the same as in the actual problem by the First Uniqueness Theorem\cite{Griffith}. This does not, however, rule out the possibility of some charge outside this cavity having to contribute to the field inside it, and so it does not prove our result.

The author has checked a dozen textbooks covering the introductory, undergraduate and graduate stages of study; no text except the one by Griffiths talks about the importance of this problem; while a few texts do talk of the superposition principle used in the second proof of this note in a different context.\cite{Reitz}$^{,}$ \cite{Purcell} 
To the best knowledge of the author, a broad discussion of this problem and the two short proofs presented here have not appeared anywhere in the literature before.


\begin{figure}[h!]
\centering
\includegraphics[scale=0.1]{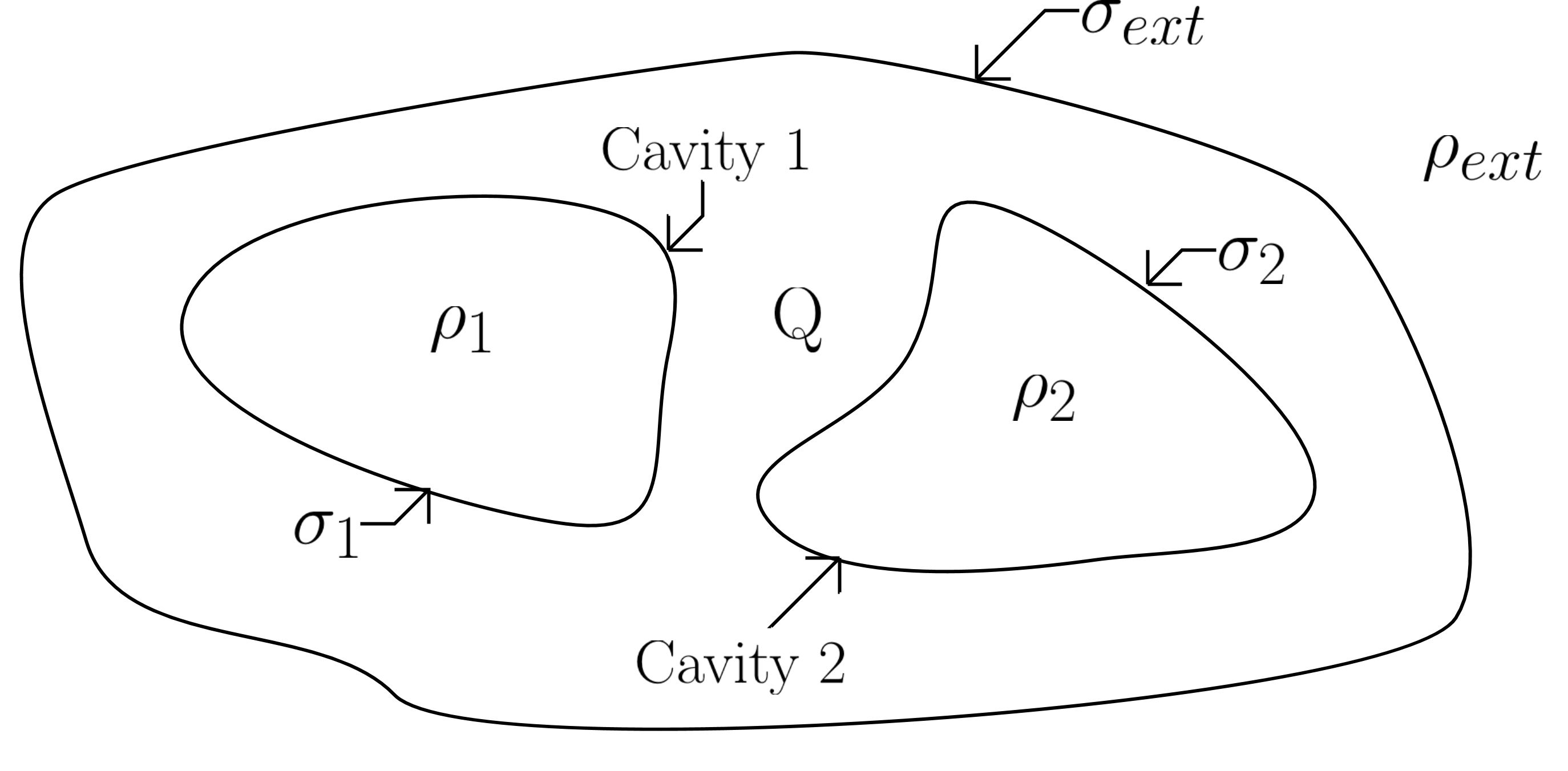}
\caption{Cross section of the conductor of arbitrary shape is shown. The volume charge densities are given by $\rho_{1}, \rho_{2},\rho_{ext}$, while $\sigma_{1}, \sigma_{2},\sigma_{ext}$ are the induced surface charge densities, in the respective domains of space. $Q$ is the given total charge on the conductor.}
\label{gasbulbdata}
\end{figure}

In our general problem, the conductor has a charge $Q$ on it. The cavity we designate as cavity 1, has a volume charge density $\rho_{1}$, while cavity 2 has volume charge density $\rho_{2}$. The exterior of the conductor has a volume charge density $\rho_{ext}$. The induced surface charge densities $\sigma_{1}$, $\sigma_{2}$, $\sigma_{3}$ are also shown in Figure 1.


\section{First Proof}

We can approach this problem by considering the constant electrostatic potential of the conductor and using the Dirichlet \cite{Jackson} uniqueness theorem, which is also noted as the First Uniqueness Theorem\cite{Griffith} in Griffiths' book. The argument in this section follows this principle. In the next section, we work directly with electric fields and do not invoke electrostatic potential. 

 The first uniqueness theorem, is stated as: \textit{The electrostatic potential inside a given volume is uniquely determined once the potential is specified on the boundary of the volume}.\cite{1}

The arguments of this proof are straightforward for the case of a conductor with one cavity. For a conductor with multiple cavities, we could inductively reduce the analysis to a conductor with fewer cavities, till we reach a conductor with one cavity. We only deal with the case of two cavities, for simplicity.

We start with the fact that the conductor is an equipotential at some constant value $V$.

First, consider the volume exterior to the outer surface of the conductor that extends to infinity. This volume goes to infinity, where the potential is zero, and is bounded by the outer surface of the conductor, where the potential is the constant $V$. By the First Uniqueness Theorem, there is a unique field distribution in this region, given the volume charge density $\rho_{ext}$ and the potentials on the boundary.  Hence we can conceive of any configuration of cavities and charges in the interior of the conductor that is different from our original configuration, specify the conductor to be at constant potential $V$; then the solution for the potential, and hence the field outside the conductor would be the same as that in our original configuration.

Armed with this freedom, we choose a configuration where we have a conductor at constant potential $V$ which has the same outer shape as our conductor, but with no cavities inside it, and thus also with no charges inside. Let's call this Configuration 1. By the First Uniqueness Theorem, the potential and the fields outside the conductor are identical in Configuration 1 and our original configuration. The fields being same, the normal components of the fields on the surface of the conductor are also the same in both configurations, and thus the induced surface charge distribution on the outer surface of the conductor are the same in either configuration. Call this surface charge density $\sigma_{ext}$.

In Configuration 1, there are no cavities and so no charges can appear in the interior of the conductor. Hence in Configuration 1, the field external to the conductor is due only to the outside charge distributions $\rho_{ext}$ and the induced $\sigma_{ext}$. By the preceding paragraph, this is true for our original configuration as well, where we have the two cavities. We have to conclude that the induced charges on the two cavity surfaces, and the volume charges $\rho_{1}$ and $\rho_{2}$ do not contribute to the fields external to the conductor in our original configuration. 

In Configuration 1, the only charges that appear are $\sigma_{ext}$ and $\rho_{ext}$. We have a solid conductor, and no electric field  can appear in the bulk of the conductor, due to these charges. These charges being the same in our original configuration, we conclude that the combined field due to charges external to the conductor, inside any of the cavities would be zero, in the original configuration.\cite{2}

It remains to show the net charge (the sum of the volume charge density and the induced surface charge density) inside cavity 1 does not produce any field inside cavity 2.\cite{3} 

For this, we consider another configuration which we designate as Configuration 2, where we have a conductor which has cavity 1, and has the external shape of the actual original conductor, but cavity 2 is covered with conducting material. This conductor is also assumed to be at the constant potential $V$. This is shown in Figure 2.

\begin{figure}[h!]
\centering
\includegraphics[scale=0.1]{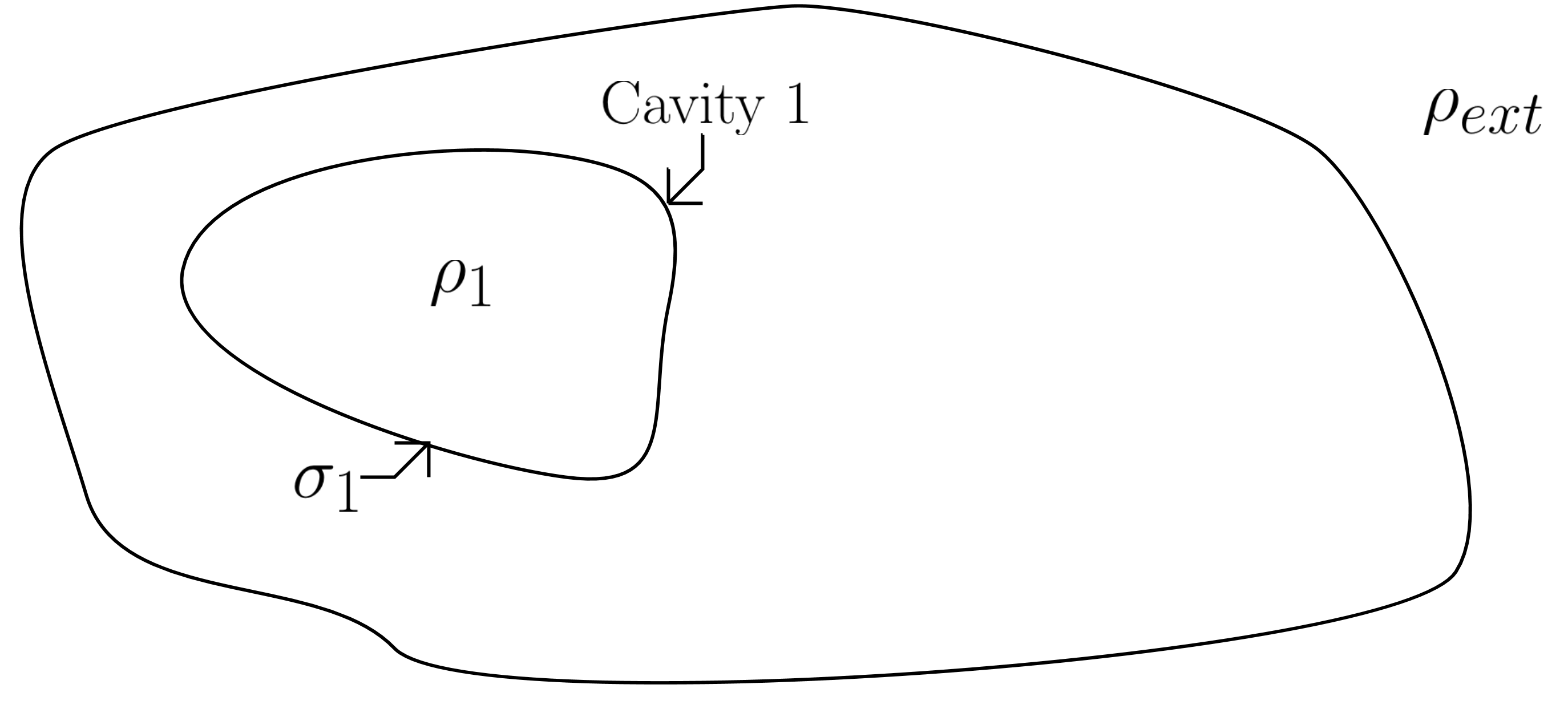}
\caption{The diagram of Configuration 2, in which cavity 2 is covered with conducting material. The volume charge density within cavity 1 and the exterior are the same as in the original configuration. The conductor is at constant potential V. }
\label{gasbulbdata}
\end{figure}

From the First Uniqueness Theorem, the field inside cavity 1 and also the induced surface charge distribution on the cavity 1 surface, would be the same in our original configuration as in Configuration 2. Call this surface charge density $\sigma_{1}$.

Now for Configuration 2, we employ the first part of the proof.\cite{2} Clearly, the field inside cavity 1 in Configuration 2 is due only to the charges $\rho_{1}$ and $\sigma_{1}$ (this is because the combined field due to volume and induced surface charges exterior to the conductor cannot penetrate into the cavity, by the first part of the proof). By the preceding paragraph, in our original configuration as well, we end up with induced charge distribution $\sigma_{1}$ in the cavity 1, and the same field distribution as in this Configuration 2. 

There are no charges in the region of cavity 2, in configuration 2, as this region is closed off with conducting material. So, the field inside cavity 1 in Configuration 2 and hence our original configuration, is due only to volume charges within cavity 1 and the surface charge induced on the surface of cavity 1. There cannot be any contribution to this field inside cavity 1, due to the charges in cavity 2. By an identical argument, the field inside cavity 2 is only due to volume and induced surface charges within cavity 2. This completes the argument.


\section{Second Proof}

We provide a second line of reasoning, using the 2nd Uniqueness Theorem as stated in Griffith's book.\cite{Griffith}. This has less stringent requirements than the Neumann boundary conditions, where the charge distribution on the conductor needs to be specified\cite{Jackson}. 

Our argument makes a simple generalisation of the free space superposition principle, to a configuration with conductors on each of which the total amount of charge has been specified.\cite{Purcell}

We reproduce the statement of the 2nd Uniqueness Theorem:  \textit{In a volume surrounded by conductors and containing a specified volume charge density $\rho$, the electric field in equilibrium is uniquely determined if the total charge on each conductor is given.}

The next result, which we use, follows from this preceding theorem.

\textit{For a configuration of conductors, when we have a charge density $\rho_{1}$, and $i$'th conductor  total charge $Q_{1_{i}}$ in electrostatic equilibrium, call the resulting field as $\vec{E}_{1}$. Similarly, when we have a charge density $\rho_{2}$ and $i$'th conductor at total charge $Q_{2_{i}}$ in electrostatic equilibrium, then call the resulting field distribution as $\vec{E}_{2}$. Then the electrostatic equilibrium field distribution for the case when the volume charge density is $\rho_{1}+\rho_{2}$ and the charge on the $i$'th conductor is $Q_{1_{i}}+Q_{2_{i}}$, is given by $\vec{E}_{1}+\vec{E}_{2}$. }

This is a straightforward and powerful result, but rarely appears in this form in many arguments in electrostatics.

Recall that originally we talked about an isolated conductor that had a total charge $Q$ on it, along with volume charges in the three separate domains.

We break up our problem in three different parts, consider te electric field due to each part, and superimpose them to get our final result. This procedure works because of the result we stated above.

We proceed as follows:

\begin{itemize}

\item In Case 1, consider a configuration in which cavity 1 has the volume charge density $\rho_{1}$, cavity 2 and the exterior of the conductor are empty, and the conductor has total charge $-q_{1}=-\int_{cavity_{1}} \rho_{1} d^{3}x$, that is, the negative of the total charge inside cavity 1.

By Gauss's theorem, we would have the charge $-q_{1}$ of the conductor distributed on the surface of the cavity 1, to make the net charge enclosing a Gaussian Volume enclosing cavity 1 to be zero.

Because of Gauss's Law, there can be no net charge on the surface of cavity 2 in this case, as cavity 2 itself does not contain any volume charge density. There can be no fields within the cavity.\cite{Feynman} The standard argument for this is given in the Feynman lectures, and in Griffiths's book, where use is made of the fact that $\nabla \times \textbf{E}=0$. In particular this implies the normal component of the electric field lines on the surface of this cavity is zero, and so there is no charge induced anywhere on the surface of cavity 2. 




As noted in the Feynman lectures\cite{Feynman}, a similar statement holds for the exterior of the conductor. There are three possibilities in this case: i)There exists a field line that begins somewhere on the outer surface of the conductor, and ends somewhere on the conductor, in which case we have a contradiction akin to the case of cavity 2; or ii) All field lines in the exterior of the conductor begin(end) on the conductor surface and and go to(come from) infinity. This again leads to a contradiction as we would have a net positive(negative) charge on the outer surface of the conductor.  

So we proved we have no fields in the exterior region, nor inside cavity 2. There is no charge distribution on the exterior surface, nor on the surface of cavity 2. We only have electric fields inside cavity 1, and that can only be due to the volume charge $\rho_{1}$ and the charge induced on the surface of cavity 1 itself. 

We note that in this case, no field lines extend from the conductor surface to infinity, and so the conductor is at the same potential as infinity, which means it is at 0 potential.

\item In the second case, consider cavity 1 and the exterior of the conductor to be empty , cavity 2 having volume charge density $\rho_{2}$, and the conductor with total charge $-q_{2}=-\int_{cavity_{2}} \rho_{2} d^{3}x$, that is, the negative of the total charge inside cavity 2.

As in the first case, we would have no charge separation in the exterior surface of the conductor nor on the surface of cavity 1, and the field inside cavity 2 would only be due to $\rho_{2}$ and the charge induced on the surface of cavity 2. No fields would be present within cavity 1 or in the exterior of the conductor.

Again the conductor is at 0 potential, as in the previous case.

\item In the third case, consider both the cavities to be empty, with the exterior of the conductor having the volume charge density $\rho_{ext}$, and the conductor having a total charge of $Q+q_{1}+q_{2}$. 

By similar arguments as before, there would be no charge separation on the surfaces of either cavity, and no fields will appear in these cavities. The total charge $Q+q_{1}+q_{2}$ would get distributed on the exterior surface of the conductor, and we would have fields in the exterior, due only to $\rho_{3}$ and the surface charge density on the external surface of the conductor.   

\end{itemize}

We now consider the superposition of the electric fields from the three cases listed above, in all the three simply connected regions. Call this superposed field $\textbf{E}$.

This superposed field $\textbf{E}$ corresponds to the case in which we have charge densities $\rho_{1}$ in cavity 1, density $\rho_{2}$ in cavity 2, and $\rho_{ext}$ in the exterior of the conductor, and total charge $Q+q_{1}+q_{2}-q_{1}-q_{2}=Q$ on the surface of the conductor. The sum total of the preceding three cases thus prove our assertion.

(Because of the 2nd Uniqueness Theorem, once we have found a solution $\textbf{E}$ corresponding to our original configuration, it must be the only unique solution.)


While we used the 2nd Uniquenesss Theorem here, we could equally well have used the 1st Uniqueness Theorem and this strategy of superposition.

\section{conclusion}

The author would be happy to see more correct and rigorous proofs of this problem, if possible. 

It is hoped the arguments of this note would help in the correct and thorough presentation of this problem to undergraduates. Many of them mistakenly think the problem is much simpler than it actually is. It should also be clear that even though high school Physics students would easily understand this problem, the majority of them would not know enough Physics to rigorously come to the correct conclusion.

\section{acknowledgement}

The author acknowledges a useful discussion with Vivek Lohani, which led the author to the formulation of the first proof presented here. The author is also grateful for feedback and helpful suggestions from Dr Henry Greenside, Dr David Griffiths, and Dr Carl Mungan.

\end{document}